\documentclass[sigconf]{acmart}

\usepackage{booktabs} % For formal tables
\usepackage[inline]{enumitem}
\usepackage{framed}
\usepackage[T1]{fontenc}

\newcommand{\rqone}{Which types of communication channels are used in open-source projects?}

\newcommand{\highlight}[1]{\begin{leftbar}\noindent \emph{#1}\end{leftbar}}

\begin{document}

\title{Poster: Communication in Open-Source Projects--End of the E-mail Era?}
%\titlenote{Produces the permission block, and
%  copyright information}
%\subtitle{}
%\subtitlenote{The full version of the author's guide is available as
%  \texttt{acmart.pdf} document}

\author{Verena K\"afer}
\orcid{0000-0002-7070-4519}
\author{Daniel Graziotin}
\orcid{0000-0002-7070-4519}
\author{Ivan Bogicevic}
\orcid{0000-0002-7070-4519}
\author{Stefan Wagner}
\orcid{0000-0002-7070-4519}
\author{Jasmin Ramadani}
\orcid{0000-0002-7070-4519}
\email{{verena.kaefer | daniel.graziotin | ivan.bogicevic | stefan.wagner | jasmin.ramadani}@informatik.uni-stuttgart.de}
\affiliation{%
    \institution{Institute for Software Technology, University of Stuttgart}
    \streetaddress{}
    \city{Stuttgart} 
    \state{Germany} 
 }

% The default list of authors is too long for headers}
\renewcommand{\shortauthors}{V. K\"afer et al.}

\begin{abstract}

Communication is essential in software engineering.
Especially in distributed open-source teams, 
communication needs to be supported by channels including mailing lists, 
forums, issue trackers, and chat systems.
Yet, we do not have a clear understanding of which communication channels stakeholders in open-source projects use.
In this study, we fill the knowledge gap by investigating a statistically representative sample of 400 GitHub projects. 
We discover the used communication channels by regular expressions on project data.
We show that (1) half of the GitHub projects use observable communication channels; 
(2) GitHub Issues, e-mail addresses, and the modern chat system Gitter are the most common channels;
(3) mailing lists are only in place five and have a lower market share than all modern chat systems combined.

\end{abstract}

%
% The code below should be generated by the tool at
% http://dl.acm.org/ccs.cfm
% Please copy and paste the code instead of the example below. 
%



\copyrightyear{2018}
\acmYear{2018}
\setcopyright{rightsretained}
\acmConference[ICSE '18 Companion]{40th International Conference on Software Engineering Companion}{May 27-June 3, 2018}{Gothenburg, Sweden}
\acmBooktitle{ICSE '18 Companion: 40th International Conference on Software Engineering Companion, May 27-June 3, 2018, Gothenburg, Sweden}\acmDOI{10.1145/3183440.3194951}
\acmISBN{978-1-4503-5663-3/18/05}

\maketitle

%!TEX root = main.tex
\section{Introduction and Related Work}

In open-source software (OSS) projects,
which are characterized by distributed environments,
a good communication system is of particular importance.
Team members have to coordinate work,
discuss tasks, solve problems, make decisions, 
and manage their projects continuously, over different time zones~\cite{bird2008}.

OSS stakeholders use several \emph{types of communication channels}, i.e., 
electronic ways of communication such as mailing lists, forums, issue trackers, and chat systems
as a practical way of exchanging information over large distances and coordinating work~\cite{guzzi2013}.

Studies show that mailing lists are the heart of any project communication~\cite{shihab2010},
thus they should be the center of attention, but discussion on development aspects is shifting to the source code repository's issue system~\cite{guzzi2013} or modern e-mail replacement systems such as Slack and Gitter~\cite{lin2016}.

There is work about which communication channels software developers use and which of them they think are important. Storey et al.~\cite{storey} collected data from a survey of 1449 developers on what communication channels they use, 
which are the most important ones, and what challenges they face. Their results show which channels developers use for which development activity, e.g., to find answers. However, the results only show which channel is used for which activity but not the distribution of channels between repositories and other activities that might be done by other stakeholders.

We identified a gap in the current understanding of which tools are effectively employed by OSS projects overall.
In this paper, we report an automated census that let us estimate the population of communication channels among GitHub software projects.

\textbf{Research Question}: \textit{\rqone}
%!TEX root = main.tex
%!TEX root = main.tex
\section{Method}

To provide a meaningful census of communication tools and exchanged information, 
we defined our population as those GitHub-hosted projects with the following characteristics:
(1) active in the last six months (January 2017 to June 2017), 
(2) having at least 20 code commits, 
and (3) having at least 10 contributors.

We gathered the population by querying the \textit{GitHub Archive}, 
thus obtaining $13,757,509$ active projects after removing duplicates.
We randomly sampled the population to check against our inclusion criteria,
and we set a sample size of $n=400$ random projects for an error margin to $5\%$ 
and confidence level to $95\%$.

Then, we developed a Python tool using a series of regular expressions for mining the communication channels of the \textit{description artifacts} 
(projects' description, README and Wiki files). 
The tool uses the GitHub APIs for querying the description artifacts, and it runs a series of regular expressions to identify the communication channels.
The expressions match how external channels are referenced in the description artifacts 
(e.g., IRC could be mentioned in ways including \textit{IRC: \#<channel>} or \textit{irc.freenode.net: \#<channel>}).

We followed a systematic process of randomly sampling $n=40$ projects for developing and testing 
our regular expressions.
We compared the tool classifications with those of a human rater
and repeated the cycle until reaching a threshold of $86\%$ of channels with respect to those found by a human, 
thus yielding a $14\%$ false negative rate. 
Furthermore, the tool had a $18\%$ false positive rate by flagging channels that were not actually channels.
%!TEX root = main.tex
\section{Results}
\label{sec:results}

The tool identified $187$ projects ($46.7\%$ of the $400$ projects) using $290$ communication channels
(min=$1.0$, 1\textsuperscript{st} quartile=$1.0$, median=$1.0$, mean=$1.15$, standard deviation=$0.96$, 3\textsuperscript{rd} quartile=$2.0$, max=$6.0$).

\highlight{\textbf{Estimation:} Less than half of the projects on GitHub employ communication channels.}

\begin{figure}[t!]
	\center
	\includegraphics[width=\columnwidth]{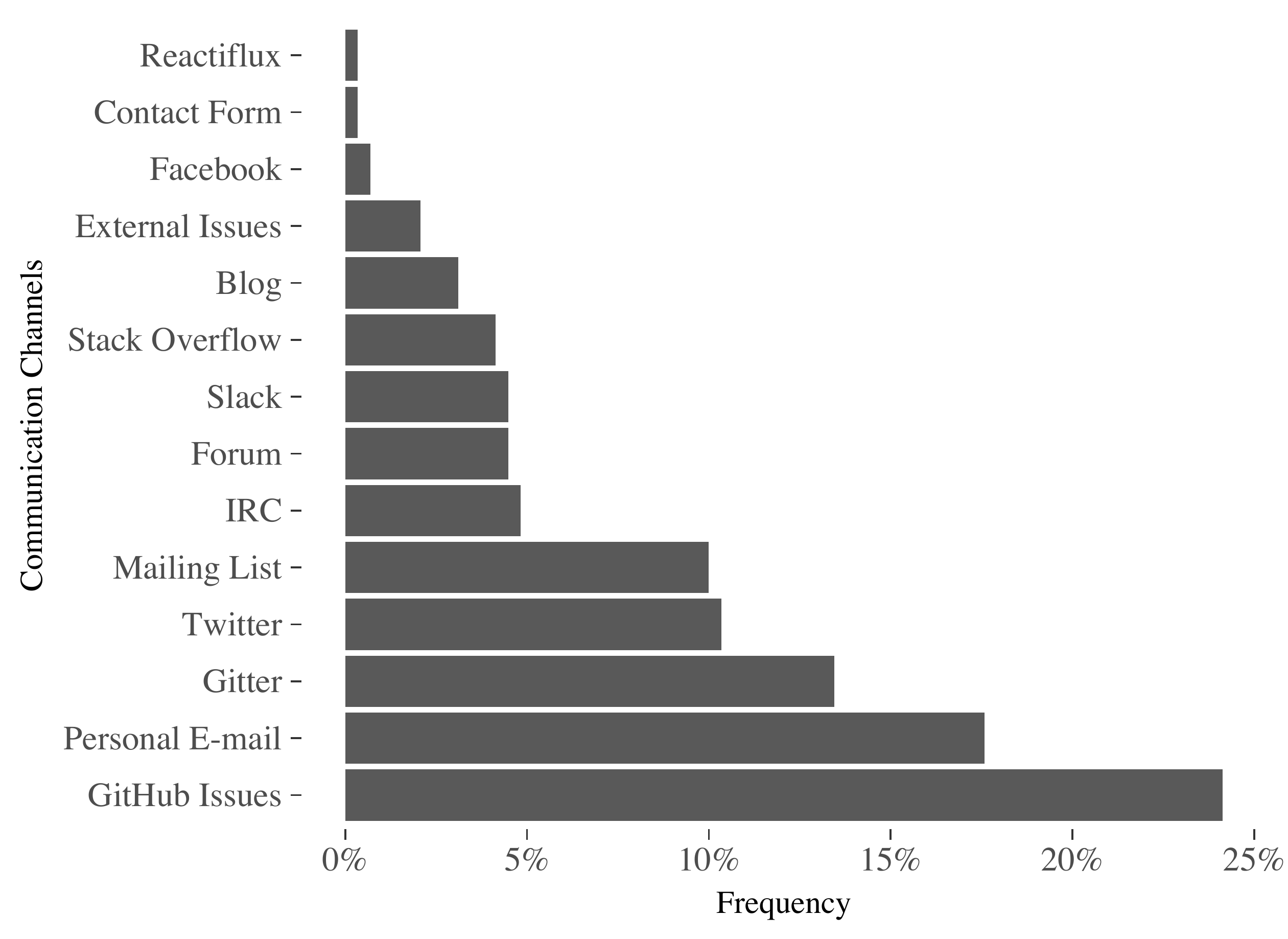}
	\caption{Types of communication channels of all communication channels in GitHub projects}
	\label{fig:channels}
\end{figure}

Figure~\ref{fig:channels} shows the identified channels and their usage.

We found the following communication channels, in order of frequency: GitHub Issues ($24.1\%$
\footnote{All percentages are relative to the $n=290$ of communication channels}
), personal e-mail ($17.6\%$), Gitter ($13.4\%$), Twitter, mailing lists ($10\%$), IRC, forums, Slack, Stack Overflow, blogs, external issue trackers, Facebook, contact forms, and Reactiflux.

Gitter usage was far higher than expected (used in $9.7\%$ of the $n=400$ GitHub projects), 
surpassing the use of mailing lists (used in $7.2\%$ of the $n=400$ GitHub projects).

Regarding the foreseen but unconfirmed decline of mailing lists~\cite{lin2016,guzzi2013,shihab2010}, 
we observe that modern e-mail replacement chat systems such as Gitter, 
Slack, and Reactiflux are replacing the use of mailing lists, 
with a combined active usage of $18.2\%$ of communication instances 
(and used in $13.25\%$ of the $n=400$ GitHub projects) 
versus the $10\%$ usage of mailing lists (and used in $7.2\%$ of the $n=400$ GitHub projects).

\highlight{\textbf{Estimation:} Mailing lists are being replaced by modern enterprise chat systems 
in OSS development.}

%\input{discussion}
%!TEX root = main.tex
\section{Discussion and Conclusion}
\label{sec:conclusion}

In this study, we performed an automatized analysis of a representative sample of open-source projects for
describing the current communication channels in use.

We show that:
\begin{itemize}
\item Only half of the projects use externally visible communication channels.
\item GitHub Issues, personal e-mail, Gitter, Twitter, and mailing lists are the most popular channels, in that order.
\item Mailing lists appear to be losing market share in favor of modern, enterprise chat systems such as Gitter and Slack.
\end{itemize}

We suggest future work to seek support for our estimation that half of all open-source projects do not employ observable
communication channels
and to understand if their omission has an effect on the efficiency or success of the projects.

Finally, we see that mailing lists are likely to disappear over time in favor of modern chat replacements,
which have a combined usage of $18.2\%$ among communication channels in GitHub projects
(estimated adoption in $13.25\%$ of projects).
Studies on the communication in open-source projects should not focus on mailing lists only but need to take the
diversity of communication channels into account.

More specifically, we are adding to the growing evidence that mailing lists are diminishing in favor of enterprise chat systems such as Gitter, Slack, and Reactiflux.

Our results are in line with the recent suggestions~\cite{lin2016} 
that enterprise chats are playing an increasingly significant role in software engineering. 
Future studies should attempt to go deep and classify the information exchanges in enterprise chat systems 
both in terms of types and frequency.

\section*{Acknowledgment}
\emph{Daniel Graziotin has been supported by the Alexander von Humboldt (AvH) Foundation.}

\bibliographystyle{ACM-Reference-Format}
\bibliography{references} 

\end{document}